\documentclass[preprint,12pt]{elsarticle}
\usepackage{amssymb,graphicx}
\usepackage[intlimits]{amsmath}
\usepackage{mathtools}
\usepackage{hyperref}
\graphicspath{{images/}}
\newcommand{\var}[1]{$\mathrm{#1}$}

\journal{Nuc. Instr. and Meth. in Phys. Res. Section A}

\begin{document}
\begin{frontmatter}
\title{On the graphical extraction of multipole mixing ratios of nuclear transitions}
\author{K. Rezynkina, A. Lopez-Martens, K. Hauschild}
\address{CSNSM, Univ. Paris-Sud, CNRS/IN2P3, Universit\'{e} Paris-Saclay, 91405 Orsay, France}
\begin{abstract}
We propose a novel graphical method for determining the mixing ratios $\delta$ and their associated uncertainties for mixed nuclear transitions. It incorporates the uncertainties on both the measured and the theoretical conversion coefficients. The accuracy of the method has been studied by deriving the corresponding probability density function. The domains of applicability of the method are carefully defined.
\end{abstract}
\begin{keyword}
nuclear transition \sep mixing ratio \sep multipolarity \sep excited states \sep gamma-ray transitions \sep internal conversion
\end{keyword}
\end{frontmatter}

\section{Introduction}
One of the tasks most often arising in nuclear spectroscopy is establishing the level scheme of a nucleus, i.e. assigning the sequence, spins and parities of the excited states of a nucleus. Conversion coefficients are valuable observables to achieve this goal. As the conversion coefficients of a transition can be calculated for various multipolarities~\cite{Kibedi} they provide important selection criteria on the spin and parity of the initial and final states. The conversion coefficients $\alpha$ can be derived through direct measurement of the $\gamma$-ray and internal conversion electrons (ICE) intensity ratios ($I_\gamma$ and $I_{ICE}$ respectively), where the total conversion coefficient
\begin{equation}
\alpha_{tot} = \frac{I_{ICE}}{I_\gamma} = \alpha_{K}+\alpha_{L}+\alpha_{M}+... ,
\end{equation}
and $\alpha_K=\frac{I_{K_{ICE}}}{I_\gamma}$, $\alpha_{L} = \frac{I_{L_{ICE}}}{I_\gamma}$, etc. are the K-shell, L-shell etc. conversion coefficients.
When the ICE measurement is missing, or incomplete, the coefficients may also be determined through the X-rays that are emitted by the atom replacing the ICE in the shells:
\begin{equation}
\alpha = \frac{I_X}{I_{\gamma} \cdot \omega },
\end{equation}
where $I_X$ is the measured intensity of the X-rays and $\omega$ is the fluorescence yield, representing the probability to fill a vacancy in the atomic shell by radiative processes. This method is limited by the measurement  thresholds in the counters used to detect the X-rays and is therefore typically restricted to K- and L-conversion.\\
 
In many cases nuclear electromagnetic transitions may have mixed multipolarities $\sigma L$, where $\sigma$ signifies the nature of the transition (electric or magnetic), and $L$ is the multipole order. The mixing ratio $\delta$ is defined as follows:
\begin{equation}
\delta^2 = \frac{I_{\gamma}'(\sigma' L')}{I_{\gamma}(\sigma L)},
\end{equation}
where $I_{\gamma}(\sigma L)$ and $I_{\gamma}'(\sigma' L')$ are the $\gamma$-ray intensities via the two mixed multipolarities $\sigma L$ and $\sigma' L'$, $L'$$>$$L$. The mixing ratio is crucial for the calculation of the experimental transition strength $B(\sigma L)$ which is an important link between theory and experiment in nuclear structure physics. Thus, the technique of determining the mixing ratios from the measured conversion coefficient is of interest to the nuclear structure physicists.\\

The relation of the $\delta^2$ to the experimental value of the conversion coefficient $\alpha_{exp}$ can be expressed as follows (see e.g.~\cite{Kantele}):
\begin{equation}
\alpha_{exp} = \frac{\alpha(\sigma L) + \delta^2 \cdot \alpha(\sigma' L')}{1 +\delta^2}.
\label{eq4}
\end{equation}
Therefore, $\delta$ equals
\begin{equation}
\delta = \sqrt{\frac{\alpha(\sigma L) - \alpha_{exp}}{\alpha_{exp} - \alpha(\sigma' L')}}.
\label{mean}
\end{equation}
One should note that determination of the mixing ratio through conversion coefficients is not sensitive to the sign of $\delta$ but only to its absolute value.
In a linear approach, the uncertainty of $\delta$ is
\begin{equation}
 \Delta \delta=\sqrt{\left(\frac{\partial \delta}{\partial \alpha(\sigma L)}\Delta \alpha(\sigma L)\right)^2 + \left(\frac{\partial \delta}{\partial \alpha(\sigma' L')}\Delta \alpha(\sigma' L')\right)^2  + \left(\frac{\partial \delta}{\partial \alpha_{exp}}\Delta \alpha_{exp}\right)^2}.
\label{ddelta}
\end{equation}
However, as it is demonstrated below, the linear approach leads to a very imprecise determination of the mixing ratio.\\

Each of the theoretical values $\alpha(\sigma L)$ and $\alpha(\sigma' L')$ has an uncertainty $\Delta \alpha$ associated to it. This uncertainty is of the order of 1-2\% and arises from two factors: the accuracy of the theoretical calculations and the accuracy of interpolation for non-tabulated values~\cite{Kibedi}.\\

In this paper we study the probability density function (PDF) of the mixing ratio, $P(\delta)$, and different methods to determine both the mean and the confidence interval. We discuss the components of $P(\delta)$, which arise from the theoretical values of $\alpha(\sigma L)$ and the experimentally determined value of $\alpha_{exp}$. We also propose a novel graphical method of extracting $\delta$ in a simple and illustrative way and study both the accuracy and the limits of appllicability. A similar graphical method was used in the 1960's~\cite{novak} to compare different theoretical calculations of conversion coefficients. However these comparisons did not take the uncertainties of the theoretical values into account. This uncertainty, even small, may result in a significant increase of the confidence interval of $\delta$ and therefore it has been incorporated in the present method.

\section{Probability density function of the mixing ratio}
The theoretical and experimental internal conversion coefficients have Gaussian PDFs associated to them. Each of these probability distributions will contribute to the PDF of the mixing ratio $P(\delta)$. Assuming the above parameters and their uncertainties to be uncorrelated, one can derive the partial PDFs  separately in the following manner: \\

\begin{equation}
\int_{0}^{+\infty} P_i(\delta) d\delta= \int_{0}^{+\infty} G(\alpha_i) d\alpha_i = 
\int_{0}^{+\infty} G(\alpha_i) \frac{\partial\alpha_i}{\partial\delta}d\delta,
\end{equation}

where $\alpha_i$ stands for  $ \alpha_1$=$\alpha(\sigma L)$, $ \alpha_2$=$\alpha(\sigma' L')$ or $\alpha_{exp}$ for each partial PDF $P_i(\delta)$: $P_1(\delta)$, $P_2(\delta)$ or $P_{exp}(\delta)$ respectively and $G(\alpha_i)$ is the Gaussian distribution. Then $P_i$ can be expressed as follows:\\

\begin{equation}
P_i(\delta) = G(\alpha_i) \frac{\partial\alpha_i}{\partial\delta}.\\
\label{eq7}
\end{equation}

 For $P_1$ and $P_2$ we express $\alpha_1$ and $\alpha_2$ as functions of $\delta$:
 \begin{equation}
 \alpha_1 = \alpha_{exp}\cdot(1+\delta^2)-\alpha_2\cdot\delta^2;\\
 \label{eq8}
 \end{equation}
 \begin{equation}
  \alpha_2 = \frac{1+\delta^2}{\delta^2}\cdot\alpha_{exp} - \frac{\alpha_1}{\delta^2}.
 \label{eq9}
 \end{equation}
 To calculate the partial PDF of $\alpha_1$ the other two parameters are fixed to their mean values. From equations (\ref{eq7}) and (\ref{eq8}) one can write (up to a normalisation constant):
\begin{equation}P_1(\delta) =\delta\cdot\exp\left(-\frac{\left(\delta^2(\alpha_{exp}-\alpha_2) + \alpha_{exp} - \mu\right)^2}{2\sigma^2}\right),
\end{equation}
 with $\mu = \alpha_1$, $\sigma=\Delta\alpha_1$.\\
 
In a similar fashion, from equations (\ref{eq7}) and (\ref{eq9}) the PDF for $\alpha_2$ is:
\begin{equation}P_2(\delta) = \frac{1}{\delta^3}\cdot\exp\left(-\frac{\left(\frac{1+\delta^2}{\delta^2}\cdot\alpha_{exp} - \frac{\alpha_1}{\delta^2}-\mu\right)^2}{2\sigma^2}\right),
\end{equation}
 with $\mu = \alpha_2$, $\sigma=\Delta\alpha_2$.\\
 From equations (\ref{eq7}) and (\ref{eq4}), the expression for $\alpha_{exp}$ is:
\begin{eqnarray}
P_{exp}(\delta) = \frac{\delta}{(\delta^2+1)^2}\cdot \exp\left(-\frac{\left(\frac{\alpha_1+\delta^2\alpha_2}{1+\delta^2} - \mu\right)^2}{2\sigma^2}\right),
\end{eqnarray}
 with $\mu$ = $\alpha_{exp}$ and $\sigma$ = $\Delta\alpha_{exp}$.\\
 
\begin{figure}[h!]
 \centering
  \includegraphics[width=1\linewidth,keepaspectratio=true]{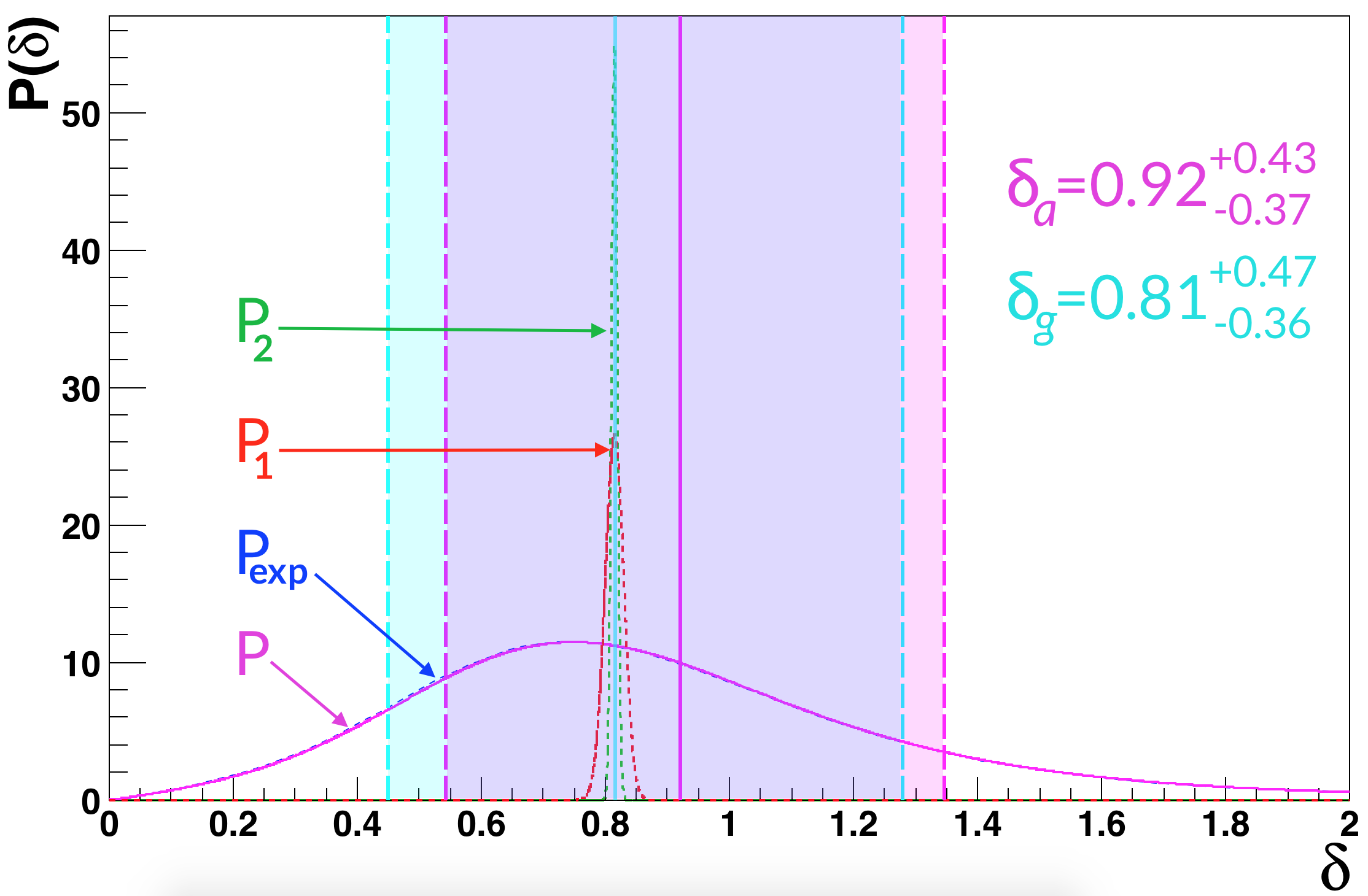}
  \caption{An example of the PDFs for the K-conversion of 200~keV transition (M2 and E3 admixture) in \var{^{251}Fm}. The dashed lines show the partial PDFs $P_1$ (red, $\alpha_K(M2)$), $P_2$ (green, $\alpha_K(E3)$) and $P_{exp}$ (blue, $\alpha_K(exp)$). The solid magenta line is the total PDF of $\delta$.  $P_1$ and  $P_2$ are normalised to 1, while $P_{exp}$ and $P$ have been normalised to 10 for purely visual reasons. The shaded cyan region represents the 68\% confidence interval around the graphically-obtained mean; the magenta shaded area is the the 68\% confidence interval around the analytically-obtained mean.}
   \label{fig:PDF}
\end{figure}

The total PDF of $\delta$ is then a convolution of these partial PDFs and therefore, in general, is no longer Gaussian:\\
 \begin{equation}
 P(\delta) = P_1(\delta) \otimes P_2(\delta)  \otimes P_{exp}(\delta).
 \label{conv}
 \end{equation}
In the following sections the method will be illustrated with a mixed M2/E3 $5/2^+\rightarrow9/2^-$ 200~keV transition in  \var{^{251}Fm}~\cite{fm}. For this particular example $\delta$ will be obtained from the K-conversion coefficient $\alpha_{K}$ measured using the GABRIELA setup~\cite{gab} installed at the focal plane of the VASSILISSA separator~\cite{shels}. Plots of the relevant PDFs calculated using RooFit~\cite{roofit} classes are shown in fig.~\ref{fig:PDF} and the corresponding parameters $\alpha_{exp}$, $\alpha_1$ and $\alpha_2$ are given in table~\ref{TAB}.\\

The mean value of $P(\delta)$ can be derived through the first order moment of the distribution
\begin{equation}
<\delta> = \int_0^{+\infty}\delta P(\delta)d\delta.
\label{eq14}
\end{equation}
The central value (from equation~\ref{eq14}) with the associated uncertainties within one standard deviation (68\% of the PDF integral calculated such that 34\% are either side of the mean value) is $\delta\,=\,0.92_{-0.37}^{+0.42}$.\\

 \begin{table}
 \centering
  \begin{tabular}{c | c | c | c }
\multicolumn{2}{c|}{}    & Mean value & Uncertainty \\ 
    \hline
$\alpha_{exp}$ & $\alpha_{K, exp}$ & 8.8 & 3.1 \\ 
    \hline
$\alpha_{1}$ & $\alpha_K(M2)$ \cite{bricc}& 14.49 & 0.21 \\ 
     \hline
$\alpha_{2}$ & $\alpha_K(E3)$ \cite{bricc}& 0.227 & 0.004 \\ 
     \hline
     \hline
& $\delta_{a}$ & 0.92& $^{+0.42}_{-0.37}$\\ 
     \hline
& $\delta_{g}$ & 0.81 & $^{+0.47}_{-0.36}$ 
  \end{tabular}
    \caption{The $\alpha_{K}$ parameters for $5/2^+\,\rightarrow\,9/2^-$ 200~keV transition in \var{^{251}Fm} used for the demonstration and the obtained analytical $\delta_a$ and graphical $\delta_g$  mixing ratio values.}
    \label{TAB}
\end{table}
   
\section{Graphical method}
We propose to trace the theoretical internal conversion coefficient $\alpha$ as a function of $\delta$ to determine both the admixture and the corresponding asymmetric uncertainty. \\

The mixed 200~keV M2/E3 transition in \var{^{251}Fm} is again used as an example. For this case the expression of the theoretical K-conversion coefficient as a function of $\delta$ (see equation~(\ref{eq4})) becomes:
\begin{equation}
\alpha_{K}(\delta) = \frac{\alpha_K(M2) + \delta^2\cdot \alpha_K(E3)}{1+\delta^2}.
\label{EQ}
\end{equation} 

\begin{figure}[h!]
 \centering
  \includegraphics[width=1\linewidth,keepaspectratio=true]{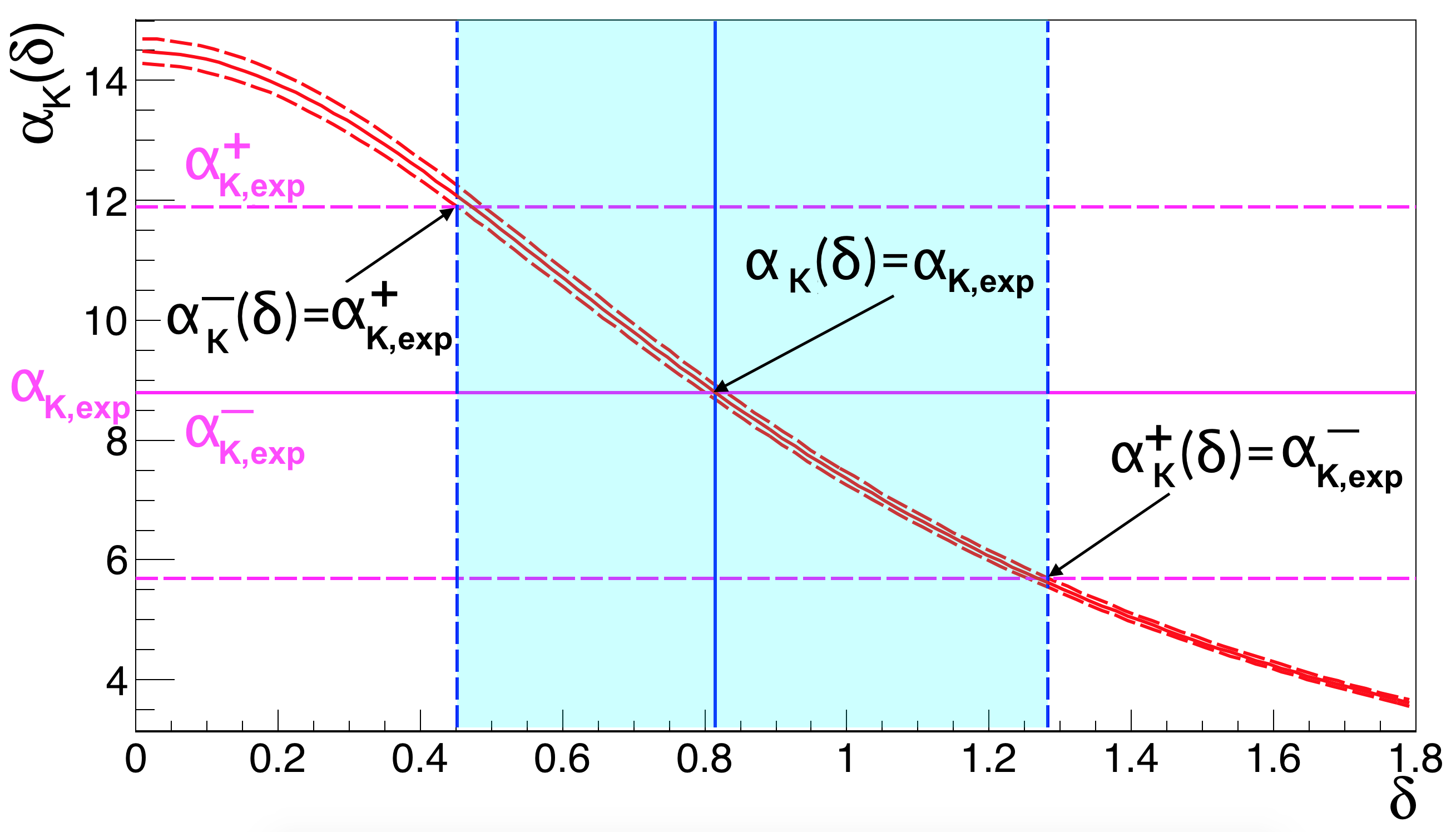}
  \caption{In red: K-conversion coefficient $\alpha_{K}$ as a function of $\delta$ with it's uncertainties; in magenta: measured value of the K-conversion coefficient $\alpha_{K, exp}$ with it's uncertainties; in blue: the deduced value of $\delta$ with the associated asymmetric uncertainties}
   \label{fig:atot}
\end{figure}

The upper and lower uncertainty limits $\alpha_{K}^\pm(\delta)$ are given by 

\begin{equation}
\begin{split}
\alpha_{K}^\pm(\delta) = \frac{\alpha_K(M2) + \delta^2\cdot \alpha_K(E3)}{1+\delta^2} \pm \\ \frac{ \sqrt{\Delta\alpha_K(M2)^2+(\delta^2\Delta\alpha_K(E3))^2}}{1+\delta^2}.
\end{split}
\end{equation} 
These theoretical curves are presented in fig.~\ref{fig:atot}.\\

The measured value of the conversion coefficient is $\alpha_{K, exp}$ with the upper and lower limits $\alpha_{K, exp}^+$ and $\alpha_{K, exp}^-$ defining the confidence interval. The central value of $\delta$ is then the solution of equation~(\ref{EQ}) with $\alpha_{K}(\delta)=\alpha_{exp}$ which is strictly identical to equation~(\ref{eq4}). \\

One should bear in mind, that the function $\alpha(\delta)$ may be decreasing as a function of $\delta$ (as is the case for $\alpha_K$ used in this example) as well as increasing (e.g. $\alpha_L$ for this same transition). Assuming the maximal error approach, the minimal value of $\delta$ is determined by the lowest intersection of the uncertainties of $\alpha_K$ and $\alpha_{exp}$, and the upper limit is the value of $\delta$ at the highest intersection.\\

The graphical method gives $\delta_g$~=~$0.81^{+0.47}_{-0.36}$ for this example. This graphical result is compared to the analytical one $\delta_a$~=~$0.92^{+0.42}_{-0.37}$ in fig.~\ref{fig:PDF}.\\

\section{Discussion}

As the theoretical uncertainties are restrained to $\sim$2\%, the error bars on $\alpha_{K, exp}$ are the main source of the uncertainty on $\delta$. It is important to note that the mean value of $P(\delta)$  (see equation~(\ref{eq14})) is, in general, not equal to the solution of equation~(\ref{mean}). For example, for the 200~keV transition in \var{^{251}Fm}, the means $<\delta>_a=0.92$ and $<\delta>_g=0.81$ are obtained using the analytical and graphical methods respectively. However, figure~\ref{fig:PDF} clearly illustrates that the solution to equation~\ref{mean} underestimates the mean value by 12\%.\\

\begin{figure}[h]
 \centering
  \includegraphics[width=1\linewidth,keepaspectratio=true]{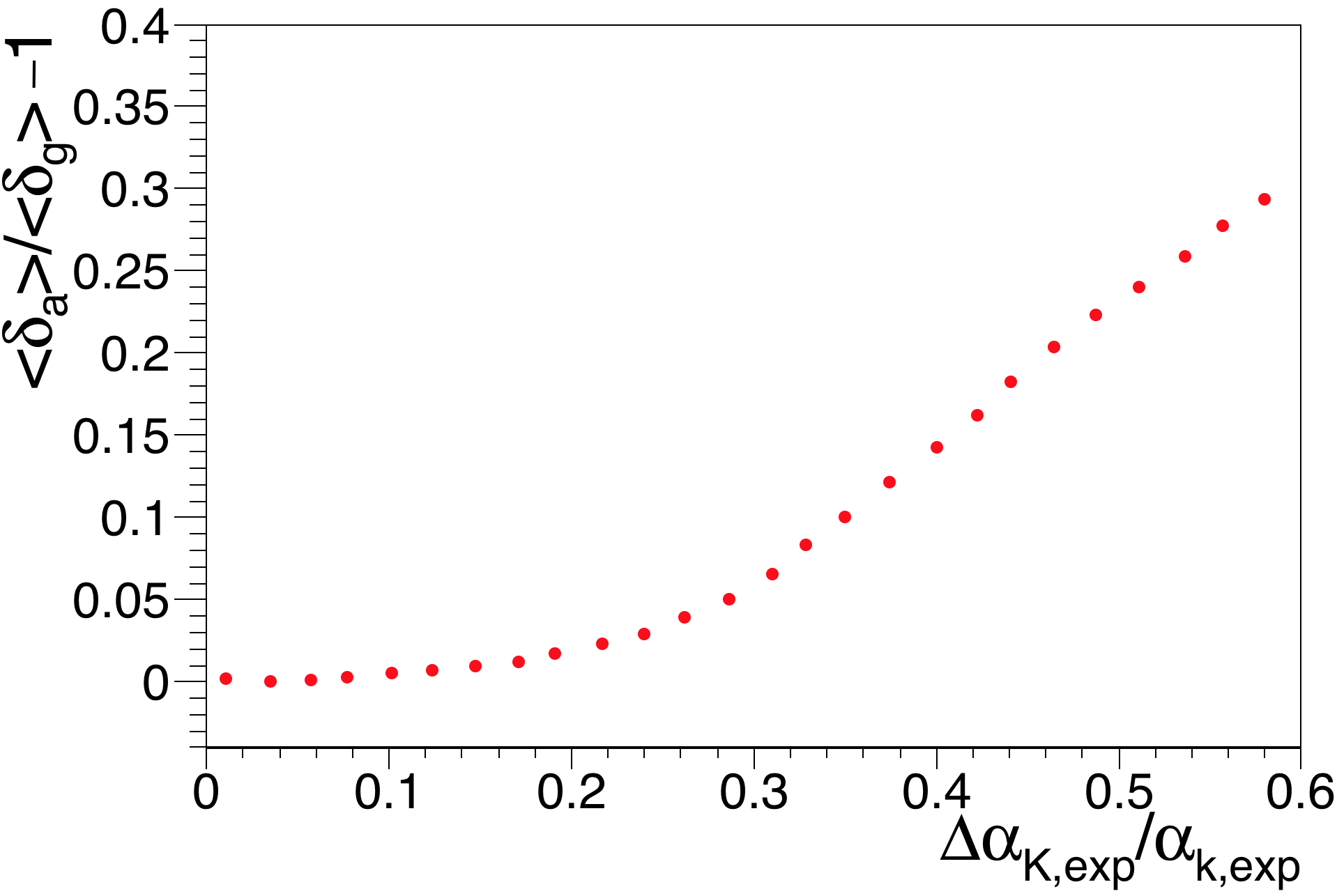}
  \caption{The dependence of the relative difference of $\delta_a$ and $\delta_g$ as a function of $\Delta\alpha_{K, exp} / \alpha_{K, exp}$.}
   \label{fig:shift}      
\end{figure}

When the uncertainty on $\alpha_{K, exp}$ is small, both formulae give the same result. This is illustrated in fig.~\ref{fig:shift} which shows how the expression $\frac{<\delta>_a}{<\delta>_g}-1$ varies as a function of the relative experimental error  $\Delta\alpha_{K, exp} / \alpha_{K, exp}$. When the relative error on $\alpha$ exceeds 25\% the mean value obtained from equation~\ref{mean} begins to deviate from the true mean value. This mismatch reaches 36\% for a relative error of 60\%.\\

The confidence interval for the of the 200~keV transition in \var{^{251}Fm} obtained with the graphical method is 6\% larger than the one obtained analytically. The comparison of the two intervals is given in fig.~\ref{fig:atot}. $P(\delta)$ provides a precise measure of the uncertainty on $\delta$. The confidence intervals derived through the graphical method are always superior or equal to the true confidence intervals extracted from $P(\delta)$, as the edge values $\alpha^-(\delta)=\alpha_{exp}^+$ and $\alpha^+(\delta)=\alpha_{exp}^-$ are beyond the one standard deviation region. \\

\section{Conclusions}
The graphical method of propagation of the uncertainties allows the central value of the admixture coefficient with its errorbars to be derived in a simple and illustrative manner. It requires much less computational power than the analytical estimate, and allows to determine the asymmetric confidence interval of $\delta$. \\

For comparison, when the value of $\delta$ is calculated in a "classic" linear approach with $<\delta>$ from equation (\ref{mean}) and $\Delta \delta$ from the equation (\ref{ddelta}), the resulting confidence interval becomes $\delta_{lin}$\,=$0.81$\,$\pm$\,$0.37$, which is an underestimate of both the central value and the confidence interval, and also does not take the asymmetry of the PDF into account.\\

It is important to notice that the graphical method only gives an upper limit for the confidence interval, and may underestimate the central value of $\delta$ when the uncertainties on the experimental conversion coefficients are high. In the cases when the graphical method is not applicable, the convolution has to be applied.\\

The graphical method also helps to better understand the influence of the different parameter values and their PDFs on the final result. The demonstrated method may also be applied to the $\alpha_K/\alpha_L$, $\alpha_L/\alpha_M$ and similar measurements, which lead to even bulkier calculations if developed analytically. \\

\end{document}